\documentclass[useAMS,usenatbib]{mn2e}
\usepackage{rotating}
\usepackage{times}
\usepackage{graphicx}
\usepackage{lscape}

\begin{document}

\title[GRB 050505: A high redshift burst detected by \textit{Swift}] {GRB 050505: A high redshift burst discovered by {\it Swift\/}}
\author[C. P. Hurkett et al.]
        {C. P.\ Hurkett$^{1}$, J. P.\ Osborne$^{1}$, K.L.\ Page$^{1}$, E.\ Rol$^{1}$, M.R.\ Goad$^{1}$, P.T.\ O'Brien$^{1}$, \newauthor A. \ Beardmore$^{1}$, O. \ Godet$^{1}$, D. N. \ Burrows$^{2}$, N. R. \ Tanvir$^{3}$, A. \ Levan$^{3}$, B. Zhang$^{4}$, \newauthor D. Malesani$^{5}$, J. E. \ Hill$^{6,7}$, J. A. \ Kennea$^{2}$, R. Chapman$^{3}$, V. \ La Parola$^{8}$, M. \ Perri$^{9}$, \newauthor P. \ Romano$^{10}$ R. \ Smith$^{11}$ \& N. \ Gehrels$^{6}$. \\
$^{1}$ X-ray \& Observational Astronomy Group, Dept. of Physics \& Astronomy, University of Leicester, Leicester LE1 7RH, U.K. \\
$^{2}$ Department of Astronomy \& Astrophysics, Penn State University, University Park, PA 16802, USA.\\
$^{3}$ Centre for Astrophysics Research, University of Hertfordshire, College Lane, Hatfield, Herts AL10 9A UK.\\
$^{4}$ Department of Physics, University of Nevada, Las Vegas, NV 89154, USA.\\
$^{5}$ International School for Advanced Studies (SISSA-ISAS), via Beirut 2-4, 34014 Trieste, Italy.\\ 
$^{6}$ NASA/Goddard Space Flight Center, Greenbelt, Maryland 20771, USA. \\
$^{7}$ Universities Space Research Association, 10211 Wincopin Circle, Suite 500, Columbia, MD, 21044-3432, USA. \\
$^{8}$ INAF - Istituto di Astrofisica Spaziale e Fisica Cosmica, Sezione di Palermo, Via La Malfa 153, I-90146 Palermo, Italy. \\
$^{9}$ ASI Science Data Center, Via Galileo Galilei, I-00044 Frascati, Italy. \\
$^{10}$ INAF - Osservatorio Astronomico di Brera, Via Bianchi 46, 1-23807 Merate, Italy.\\
$^{11}$ Astrophysics Research Institute, Liverpool John Moores University, Twelve Quays House, Birkenhead CH41 1LD, UK. \\
\\}

\date{Accepted 2005 ??  ??. Received 2005 ?? ??; in original form 2005 ?? ??}

\pagerange{\pageref{firstpage}--\pageref{lastpage}} \pubyear{2005}

\maketitle

\label{firstpage}

\begin{abstract}
We report the discovery and subsequent multi-wavelength afterglow behaviour of the 
high redshift ($z = 4.27$) Gamma Ray Burst GRB 050505. This burst is the third most distant burst, measured by spectroscopic 
redshift, discovered after GRB 000131 ($z = 4.50$) and GRB 050904 ($z = 6.29$). GRB 050505 is a long GRB with a multipeaked 
$\gamma$-ray light curve, with a duration of $T_{90} = 63\pm2$ s and an inferred isotropic release in $\gamma$-rays of 
$\sim4.44\times 10^{53}$ ergs in the $1-10^{4}$ keV rest frame energy range. The {\it Swift\/} 
X-Ray Telescope followed the afterglow for $14$ days, detecting two breaks in the light curve at $\mathrm{7.4}^{+1.5}_{-1.5}$ ks 
and $\mathrm{58.0}^{+9.9}_{-15.4}$ ks after the burst trigger. The power law decay slopes before, between and after these 
breaks were $0.25^{+0.16}_{-0.17}$, $1.17^{+0.08}_{-0.09}$ and $1.97^{+0.27}_{-0.28}$ respectively. The light curve 
can also be fit with a `smoothly broken' power law model with a break observed at $\sim T+18.5$ ks, with decay slopes of 
$\sim0.4$ and $\sim1.8$ before and after the break respectively. The X-ray afterglow shows no spectral variation over the 
course of the {\it Swift\/} observations, being well fit with a single power law of photon index $\sim1.90$. This behaviour 
is expected for the cessation of continued energisation of the ISM shock followed by a break caused by a jet, either uniform 
or structured. Neither break is consistent with a cooling break. The spectral energy distribution indeed shows the cooling 
frequency to be below the X-ray but above optical frequencies. The optical -- X-ray spectrum also shows that there is 
significant X-ray absorption in excess of that due to our Galaxy but very little optical/UV extinction, with $E(B-V)$ 
$\approx0.10$ for a SMC-like extinction curve. 

\end{abstract}


\section{Introduction}

Gamma Ray Bursts (GRBs) are expected to be visible over a large range of redshifts with a potential upper limit of $z 
\sim15-20$~\cite{Lamb_2000}. The lowest recorded GRB redshift to date is GRB 980425 with 
$z = 0.0085\pm0.0002$~\cite{Tinney_1998}, whilst the highest is GRB 050904 at $z=6.29\pm0.01$~\cite{Kawai_3937}\footnote{We 
also note that a photometric redshift of $\sim6.6$ has been reported for GRB 060116 (Malesani et al. 2006,
Grazian et al. 2006) but this has yet to be confirmed spectroscopically.} Bursts at high redshift are potentially 
important since they can be powerful probes of the early Universe. Long duration bursts ($T_{90} \geq 2$ s) are 
likely caused by the core-collapse of a massive star~\cite{Hjorth_2003,Stanek_2003}, linking these bursts directly 
to contemporary star formation. In addition, high redshift GRBs allow us to probe the intervening matter between the 
observer and GRB, and particularly the conditions of their host galaxies (e.g. Vreeswijk et al. 2004).

So far, only $\sim50$ bursts have a firm redshift determination, mostly obtained through spectroscopy of their 
optical afterglow. The record holder is GRB 050904, see Watson et al. (2005), Cusumano et al. (2005) and 
Tagliaferri et al. (2005) for more details. Previously the highest redshift burst was GRB 000131~\cite{Andersen_2000}. 
Unfortunately {\it BATSE\/} detected GRB 000131 during a partial data gap~\cite{Kippen_2000} so its position was not 
localised until $56$ hours after the trigger, thus its early time behaviour is unknown. No breaks were directly observed 
in the light curve for GRB 000131 but, based on the spectral index, an upper limit on the jet break time of $<3.5$ 
days has been hypothesised ~\cite{Andersen_2000}. In contrast, the rapid position dissemination for GRB 050505 allowed 
a rapid redshift determination, and its automated follow-up program provided a well-covered X-ray afterglow light curve. 
Here we present the results from {\it Swift\/}~\cite{Gehrels_2004} on GRB 050505.
Two breaks were detected in the X-ray light curve, the first of which we consider to be due to the cessation of continued
energisation of the ISM shock and the second is a jet break, caused by either a structured or uniform jet. Both breaks are 
inconsistent with a cooling break passing through the X-ray band (see $\S4.1$).

\section{\textit{Swift} Observations of GRB 050505.}

At 23:22:21 UT on the $5^{\rm th}$ of May $2005$, the {\it Swift\/} Burst Alert Telescope (BAT; Barthelmy et al. 2005), triggered 
and located GRB 050505 on-board (trigger ID $117504$; Hurkett et al. 2005). The BAT mask-weighted light curve (see Fig $1$) 
shows a multi-peaked structure with a $T_{90}$\footnote{The time during which $90\%$ of the counts are accumulated} ($15-350$ keV) 
of $63\pm2$ seconds. The initial peak began $\sim15$ seconds before the trigger and extended to $10$ seconds after the 
trigger. There were three further short spikes with peaks at $T+22.3$, $T+30.4$ and $T+50.4$ seconds, where $T$ is the 
trigger time. 

The $T_{90}$ observed $15-150$ keV BAT spectrum was adequately fit by a single power law with a photon index $= 1.56\pm0.12$
(with $\chi^{2}/DOF = 48/56$) and a mean flux over $T_{90}$ of ($6.44^{+0.42}_{-1.54}$)$\times10^{-8}$ ergs 
cm$^{-2}$ s$^{-1}$ in the $15-350$ keV range and ($3.76^{+0.21}_{-0.69}$)$\times10^{-8}$ ergs cm$^{-2}$ s$^{-1}$ 
in the $15-150$ keV range. All errors in this paper are quoted at $90\%$ confidence unless otherwise stated. Whilst 
fitting a cutoff power law does not give a significantly better fit ($\chi^{2}/DOF = 45/55$) it does provide us 
with an indication of the $E_{\rm peak}$ for this burst. We find a photon index $= 1.02^{+0.51}_{-0.57}$ and a lower 
limit of $E_{\rm peak,obs}$ \(>\) $52$ keV (at the $90\%$ confidence level). 

The burst was detected in each of the four standard BAT energy bands and had a ratio of fluence in the $50-100$ keV 
band to that in the $25-50$ keV of $1.37\pm0.14$, close to the mean ratio of the {\it BATSE\/} 
catalogue\footnote{http://cossc.gsfc.nasa.gov/docs/cgro/batse/4Bcatalog/index.html}. The total fluence in the $15-350$ keV band is 
($4.1\pm0.4$)$\times10^{-6}$ ergs cm$^{-2}$~\cite{Hullinger}, which is slightly higher than the average fluence detected to 
date by {\it Swift\/}.

\begin{figure}
\begin{center}
\includegraphics[width=6cm,angle=270]{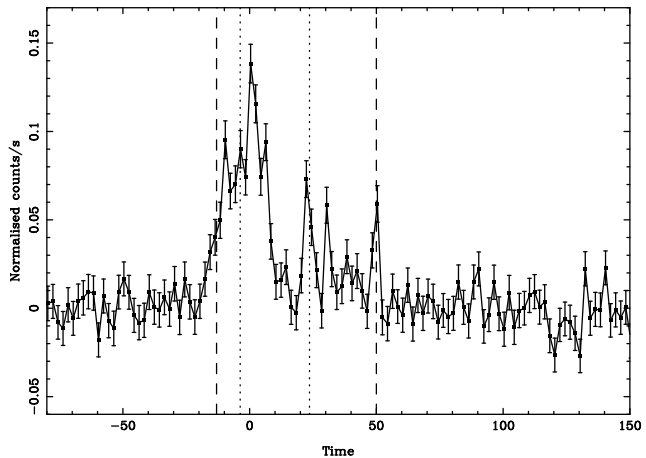}
\caption{The BAT mask weighted light curve ($15-350$ keV), where $T=0$ is the trigger time. The dashed lines indicate the $T_{90}$ interval and the dotted lines indicate the $T_{50}$ interval.}
\end{center}
\end{figure}

{\it Swift\/} executed an automated slew to the BAT position and the X-Ray Telescope (XRT; Burrows et al. 2005), began taking data at 
00:09:23 UT on $6^{\rm th}$ May 2005, $\sim47$ minutes after the burst trigger. The delay in the spacecraft slew was 
due to an Earth limb observing constraint. The XRT was in Auto state, where autonomous data mode switching was enabled, 
but the on-board software did not automatically locate a position due to low source brightness. Ground processing 
revealed an uncatalogued X-ray source within the BAT error circle located at RA = 09:27:03.2, Dec = +30:16:21.5 (J2000) 
with an estimated uncertainty of $6$ arcseconds radius ($90\%$ containment; Kennea et al. 2005). Updating the XRT boresight,
Moretti et al. (2005) have corrected this position to RA = 09:27:3.16, Dec = +30:16:22.7 with an estimated uncertainty 
of $3.2$ arcseconds (also $90\%$ containment). No data was obtained in WT mode due to the delayed slew, since this mode
is only used for sources brighter than 1 mCrab.

Observations continued over the next $14$ days, though the X-ray afterglow was not detected after the $6^{\rm th}$ day. 
Co-adding the final $8$ days of observations produced a total of $58$ ks of data providing an upper limit of 
$\sim3.5\times10^{-4}$ counts s$^{-1}$, consistent with the extrapolated decay (see $\S2.1$).

The {\it Swift\/} Ultra-Violet/Optical Telescope (UVOT; Roming et al. 2005), observed the field starting at 00:09:08 UT on the $6^{\rm th}$ 
May 2005, $\sim47$ minutes after the burst trigger. The initial data were limited to one $100$ second exposure in each 
of the four filters. No new sources were found in the XRT error circle to limiting magnitudes ($5\sigma$ in $6$ arcsecond 
radius apertures) of $\rm V>17.7$, $\rm U>18.4$, $\rm UVW1>18.9$ and $\rm UVM2>19.7$. Additional co-added, deeper exposures 
($\sim2000$ s) with the UVOT also failed to detect an optical counterpart at the location of the GRB~\cite{Rosen_2005a,Rosen_2005b}. 
The deeper exposure in V placed a limiting magnitude for the source at $>20.35$ (3$\sigma$ confidence level) for a 
total exposure of $2527$ s co-added from a series of short exposures over the time span of $2807$ s to $28543$ s after the 
trigger. Due to the delayed slew of the satellite we cannot confirm whether this burst was intrinsically subluminous or
had faded below the detection level of the UVOT. However, the optical counterpart for this burst was detected by several other 
facilities (see Table $2$), which argues for the case that it was merely too faint to be detected by the UVOT $\sim47$ minutes
post-burst.

\subsection{X-ray Light curve and Spectral Analysis.}

In the PC mode the XRT suffers from pile-up when the count-rate is $\geq 0.8$ counts s$^{-1}$~\cite{Vaughan_2005}. To counter the 
effects of pile-up we extracted a series of grade $0-12$ spectra from the first $23$ ks of data using annuli of varying inner radii. 
These background corrected spectra were then fitted in XSPEC with an absorbed power law. We deem the point at which pile-up no longer 
affects our results to be when the power law index does not vary when the inner radius of the annulus is increased. For GRB 050505 
this occurred when we excluded the inner $8$ pixels (radius). Data after $T+23$ ks were not piled up and therefore required 
no correction.

The X-ray light curve of GRB 050505 is shown in Fig $2$ and $3$, with observations starting at $T+3$ ks after the trigger time and 
extending to $T+1.05\times10^{3}$ ks. We characterise the behaviour of the XRT flux in terms of the standard power law 
indices $f \propto \nu^{-\beta}t^{-\alpha}$. Thus a series of power law models were fit to the light curve data. The simplest model 
considered was a single power law of decay index $\alpha$. This model was rejected for GRB 050505 as it gave an unacceptable value 
of $\chi^{2}/DOF = 122.5/46$.

`Broken' and `doubly broken' power laws were also fitted to the data. These models consist of two or three (respectively) 
power law sections whose slopes join but change instantaneously from $\alpha_{i}$ to $\alpha_{i+1}$ at the break times.
A `broken' power law model is also a poor description of the lightcurve ($\alpha_{1} = 0.90^{+0.05}_{-0.05}, \alpha_{2} = 
1.80^{+0.18}_{-0.15}, t_{\rm break} = 42^{+6.7}_{-7.3}$ ks) with $\chi^{2}/DOF = 58.0/44$. A `doubly broken' power law provides
a much better statistical fit to the data with $\chi^{2}/DOF = 38.7/42$ ($>99.9\%$ improvement over both the simple and the 
broken power law). The model consists of a shallow decay, $\alpha_{1}$ = $0.25^{+0.16}_{-0.17}$, which breaks sharply at 
$t_{1} = 7.4^{+1.5}_{-1.5}$ ks to a steeper decay of $\alpha_{2} = 1.17^{+0.08}_{-0.09}$. The steeper decay breaks sharply again 
at $t_{2} = 58^{+9.9}_{-15.4}$ ks into a yet more rapidly decaying index of $\alpha_{3} = 1.97^{+0.27}_{-0.28}$. 

A `smoothly broken' power law was also fit to the data, it consists of two power law sections; however, the transition between these 
slopes is not instantaneous, but may spread over several decades in time: 
\begin{eqnarray}
f(t) & = & K((\frac{t}{t_{b}})^{-\alpha_{1}S}+(\frac{t}{t_{b}})^{-\alpha_{2}S})^{1/S} , 
\end{eqnarray} 
where $S$ is the smoothing parameter, $t_{\rm b}$ is the break time and $K$ is a normalisation constant. This produces a 
smooth break rather than a sharp break as in the previous models. Typically the values of the smoothing parameter, $S$, 
reported in the literature range between $0.5-10$, with a value of $\sim1$ being favoured both observatonally and 
theoretically~\cite{Stanek_2005,Beuermann_1999}. A larger value of the smoothing parameter gives a sharper break. The 
light curve of GRB 050505 is well fit by a smoothly broken power law with $\chi^{2}/DOF \sim1.0$. Unfortunately there 
is degeneracy between the smoothing factor and the initial decay index, with any value of $S$ between $0.5$ and $3$ producing 
a good fit to the data (limit of $\chi^{2}/DOF = 1.16$). However, if we constrain the model parameters so that 
$\alpha_{1}$ must have a positive value and that $\alpha_{2}$ equals $p$, the electron spectral index (calculated from our 
spectral index, $\beta$,~\cite{Zhang_2004}), then we find that a smoothing parameter in the range of $0.5-2$ is allowed. 
This range of smoothing factors produces $\alpha_{1} \sim0.5$. Restricting $S$ to $1.0$ we find $\alpha_{1} = 
0.37^{+0.13}_{-0.15}$, $\alpha_{2} = 1.80^{+0.16}_{-0.16}$, $t_{\rm break} = 18.5^{+4.4}_{-3.2}$ ks
and $\chi^{2}/DOF = 46.9/45$ (see Fig $3$).

\begin{figure}
\begin{center}
\includegraphics[width=5.5cm,angle=-90]{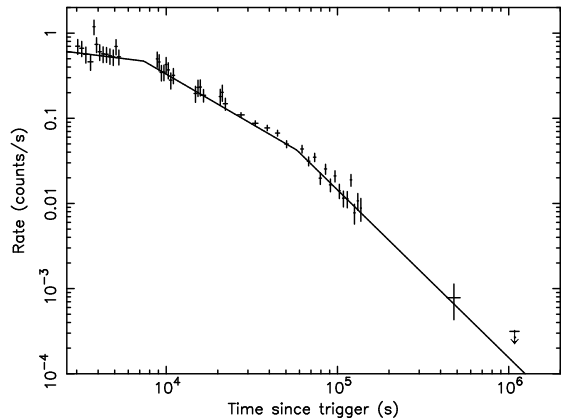}
\caption{The $0.3-10.0$ keV X-ray light curve of GRB 050505 fit to a doubly broken power law (see $\S$2.1). The first decay slope, $\alpha_{1} = 
0.25^{+0.16}_{-0.17}$, which breaks sharply at $t_{1} = 7.4^{+1.5}_{-1.5}$ ks (observer's frame) to second decay slope of 
$\alpha_{2} = 1.17^{+0.08}_{-0.09}$. A second break occurs at $t_{2} = 58^{+9.9}_{-15.4}$ ks into a third 
decay slope of $\alpha_{3} = 1.97^{+0.27}_{-0.28}$. The final point on the light curve is the $3\sigma$ upper limit to the 
detection of the afterglow at that time.}
\end{center}
\end{figure}

\begin{figure}
\begin{center}
\includegraphics[width=5.5cm,angle=-90]{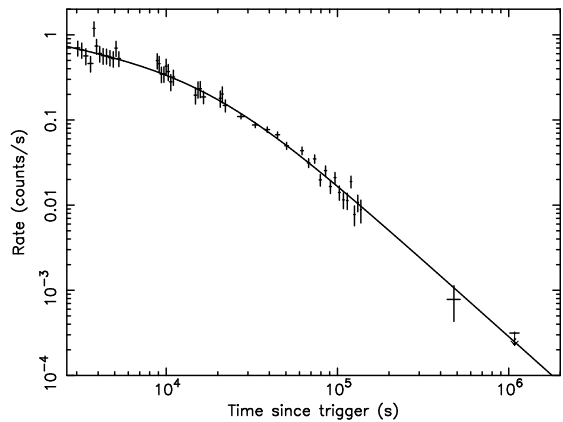}
\caption{As Fig. $2$, but fit with a smoothly broken power law (see $\S2.1$). The first decay slope, $\alpha_{1} = 0.37^{+0.13}_{-0.15}$, 
which breaks at $t = 18.5^{+4.4}_{-3.2}$ ks (observer's frame), with a smoothing parameter $S = 1.0$, 
to second decay slope of $\alpha_{2} = 1.80^{+0.16}_{-0.16}$.}
\end{center}
\end{figure}

Spectral fits were performed over $0.3-10.0$ keV using grade $0-12$ events (as selected for the light curve analysis), binned to 20
counts per data point, individually for co-added data encompassing $T+3$ to $T+17$ ks and $T+26$ to $T+138$ ks, as well as the summed 
spectra for both intervals combined (see Table $1$).

The spectra were fit with a power law model (see Fig $4$) with the absorption, $N_{\rm H}$, set at the Galactic 
column density ($2.1\times10^{20}$ cm$^{-2}$, Dickey \& Lockman 1990), and with power law models with excess absorption 
(either in our Galaxy or the GRB host galaxy). During our analysis both Wisconsin and T{\"u}bingen-Boulder ISM absorption 
models~\cite{xspec} were used; there was no 
significant difference in either the statistical quality of the fit or in the resulting derived parameters between 
the two. We present results obtained using the T{\"u}bingen-Boulder model using the local interstellar medium abundances 
of Anders and Grevesse~\cite{Anders_1989}~\footnote{We also preformed spectral fits using the abundances of Wilms, Allen and McCray~\cite{Wilms_2000} 
and found that they produced $N_{\rm H}$ values that agreed, within errors, to those given by the Anders and Grevesse abundances.}.

It is clear from Table $1$ that there is no evidence for spectral change over the duration of the observations. This
was confirmed by making a hardness ratio time series in the bands $0.3-1.5$ keV and $1.5-10.0$ keV, no variation was apparent.
The fit to the total data-set reported in Table $1$ also shows that there is significant excess absorption in this spectrum (at 
$>99.99\%$ confidence). Statistically both Galactic and extra-galactic absorption fits appear equally likely, however, 
if the excess absorption were to be due to gas in our Galaxy alone then the value of the excess absorption is almost 
twice the column density quoted by Dickey and Lockman ~\cite{Dickey_1990}. Therefore, we conclude that the bulk component 
of excess absorption must come from the host galaxy with a value of $N_{\rm H}$ = $1.28^{+0.61}_{-0.58} \times10^{22} \rm cm^{-2}$ 
assuming local ISM abundances in the GRB rest frame.

The photon index $= \beta + 1 = 1.90^{+0.08}_{-0.08}$, is typical of the photon indices seen in other GRB afterglows~\cite{Nousek_2005}, 
even though we are sampling a higher range of spectral energies due to the high redshift of this burst. 
With a redshift of $4.27$~\cite{Berger_3368} we are measuring the spectrum over a rest-frame range of $1.6-53$ keV. The spectrum is well modelled up to 
such high energies in the rest frame of the GRB, and the photon index is comparable to the values found from low redshift bursts.

\begin{table*}
\centering
   \begin{tabular}{@{}lccccccccc@{}}
   \hline
Model$^{a}$ & \multicolumn{3}{c}{Co-added data for $T+3$ - $T+17$ ks} & \multicolumn{3}{c}{Co-added data for $T+26$ - $T+138$ ks} & \multicolumn{3}{c}{All data co-added} \\ \cline{2-4} \cline{5-7} \cline{8-10}
 & Photon & Excess $N_{\rm H}$ & $\chi^{2}$ (DOF) & Photon & Excess $N_{\rm H}$ & $\chi^{2}$ (DOF) & Photon & Excess $N_{\rm H}$ & $\chi^{2}$ (DOF) \\
 & index         & ($10^{20}$ cm$^{-2}$) &                   & index        & ($10^{20}$ cm$^{-2}$) &                   & index        & ($10^{20}$ cm$^{-2}$) & \\

   \hline
PL+Gal & $1.76^{+0.09}_{-0.09}$ & - & $26.9$ (27) & $1.77^{+0.06}_{-0.06}$ & - & $86.2$ ($69$) & $1.76^{+0.05}_{-0.05}$ & - & $133$ ($97$) \\
PL+Gal+Abs & $1.91^{+0.19}_{-0.18}$ & $ <7.74$ & $24.2$ ($26$) & $1.94^{+0.12}_{-0.11}$ & $3.91^{+2.43}_{-2.14}$ & $77.3$ ($68$) & $1.93^{+0.10}_{-0.10}$ & $3.81^{+2.09}_{-1.93}$ & $102$ ($96$) \\ 
PL+Gal+ZAbs$\ast$ & $1.87^{+0.15}_{-0.14}$ & $113^{+123}_{-107}$ & $23.9$ ($26$) & $1.91^{+0.10}_{-0.09}$ & $133^{+73}_{-65}$ & $74.7$ ($68$) & $1.90^{+0.08}_{-0.08}$ & $128^{+61}_{-58}$ & $99$ ($96$) \\

$\ast$ z fixed at $4.27$ & & & & & \\
 
  \hline
   \end{tabular}
%\begin{minipage}{4cm}
%helloah
%\end{minipage}
\caption{Spectral fits for GRB 050505. The spectra show no variation. Whilst an absorbed power law is sufficient to model the data it can be seen that an additional absorption component proves a better fit, particularly at high redshift. $^{a}$ Spectral models: power-law (PL), Galactic absorption (Gal), which has been assumed to be $2.1\times10^{20}$ cm$^{-2}$ (Dickey \& Lockman 1990), excess Galactic absorption (Abs) and excess absorption in the host galaxy (ZAbs). }
\end{table*}

\section{Follow-Up Detections of GRB 050505.}

The first reported detection of an optical counterpart for GRB 050505 was made by Cenko et al.~\cite{Cenko_3366} observing from 
the Keck I telescope, quickly followed by a measurement of the redshift by the same collaboration~\cite{Berger_3368}. See Table 
$2$ for a summary of all of the optical observations reported on the GCN network as well as data from Faulkes Telescope North, 
reported here for the first time.

\begin{table*}
\centering
   \begin{tabular}{@{}llrrrrlrlr@{}}
   \hline
Filter & Limiting mag. & Detected mag. & Duration (s) & Mid-point time (s) & Observatory & References \\
   \hline
$R$ & $9.2$ & & $30$ & $0$ & BOOTES - Very Wide Field Camera & Jelinek et al. 2005 \\
$I$ & & $18.2\pm0.2$ & $456$ & $796$ & TAROT & Klotz et al. 2005 \\
$I$ & & $18.4\pm0.2$ & $456$ & $1259$ & TAROT & Klotz et al. 2005 \\
$I$ & $18.8$ & & $584$ & $1946$ & TAROT & Klotz et al. 2005 \\
$R$ & $18.5$ & & $1680$ & $2398$ & AAVSO & Hohman et al. 2005 \\
$R$ & $19.0$ & & $3600$ & $2799$ & BOOTES - IR & de Ugarte Postigo et al. 2005 \\
$R$ & $19.7$ & & $2100$ & $14326$ & SARA & Homewood et al. 2005 \\
$I$ & & $20.51\pm0.05$ &	- & 23006 & Keck & Berger et al. 2005b \\
$g$ & & $423.67\pm0.12$ &	- & 23006 & Keck & Berger et al. 2005b \\
$K$ & & $18.1\pm0.2$ & - & $24552$ & UKIRT WFCAM & Rol et al. 2005 \\
$R$ & & $21.8\pm0.1$ & $540$ & $29894$ & Faulkes Telescope North & this paper\\
$i^{\prime}$ & & $21.0\pm0.2$ &	$520$ & $30154$ & Faulkes Telescope North & this paper\\
$B$ & $21.9$ & & $640$ & $30154$ & Faulkes Telescope North & this paper\\
   \hline
   \end{tabular}
\caption{Optical follow-up of GRB 050505. Mid-point times are given in seconds after the Trigger time.}
\end{table*} 

Unfortunately the initial spacecraft message sent to the GCN network erronously flagged this event as not 
a GRB, which consequently meant that the majority of robotic follow-up missions did not observe GRB 050505 
promptly. The sparse nature of this combined data-set naturally limits the knowledge that can be obtained.

\section{Discussion}

\subsection{Physical Origin of the Light curve Break}

A doubly broken power law fit contains breaks at $7.4^{+1.5}_{-1.5}$ ks and $58.0^{+9.9}_{-15.4}$ ks in the observer's
frame, which translate to $T+1.4^{+0.3}_{-0.3}$ ks and $T+11.0^{+1.9}_{-2.9}$ ks in the rest frame of the burst. The
amplitudes of these temporal breaks are $\Delta \alpha_{1-2} = 0.92\pm0.19$ and $\Delta \alpha_{2-3} = 0.80\pm0.29$.

The combined BAT and XRT light curve (shown in Fig $5$) is consistent with the schematic diagram (fig $3$ of Nousek 
et al. 2005) of the canonical behaviour of {\it Swift\/} XRT early light curves. For GRB 050505 there is necessarily
a steep decline from the bulk of the BAT emission to the early XRT emission, which would comprise the first power law 
segment identified by Nousek et al., the early flat slope of the XRT 
decay ($\alpha_{1}$) would comprise the second segment of canonical decay and the second slope of the doubly broken 
power law fit ($\alpha_{2}$) would comprise the third canonical segment. The BAT and XRT light curves are consistent 
with joining in the $\sim47$ minute gap that separates them (see O'Brien et al. 2005), though this behaviour cannot 
be confirmed with the data we have available.

Light curve breaks can be caused by the passage through the X-ray band of the cooling frequency, the ending of continued 
shock energization, the presence of a structured jet or jet deceleration causing the relativistic beaming to become broader 
than the jet angle. We examine these possibilities here.

We can immediately rule out the presence of a cooling break for either break as this would result in $\Delta \alpha = 0.5$ 
and a change in spectral index~\cite{Sari_1998}. 

Either of the X-ray light curve breaks might represent the end of the energy injection into the 
forward shock of the relativistic outflow~\cite{Nousek_2005,Zhang_2005}, given the lack of spectral variation (and 
presuming the emission before the break was dominated by the forward shock). However, the temporal placement of the first 
break makes it the more favourable of the two for this interpretation.

\begin{figure}
\begin{center}
\includegraphics[width=5.5cm,angle=270]{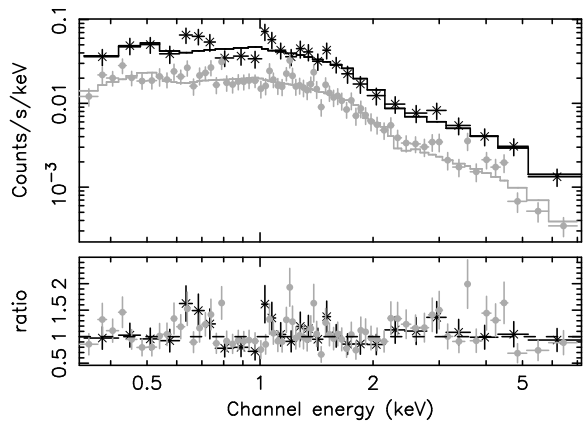}
\caption{The summed $0.3-10.0$ keV spectrum of GRB 050505 from 'piled up' (crosses) and 'non piled up' (solid circles) data, which are consistent with a photon index of $\sim1.90$, Galactic absorption ($2.1\times10^{20}$ cm$^{-2}$) plus an excess absorption component from the host galaxy ($128\times10^{20}$ cm$^{-2}$). See Table $1$ for a summary of spectral models.}
\end{center}
\end{figure}

Nousek et al. (2005) consider that a shallow flux decay is caused by continuous energy injection into the forward 
shock either due to a decrease in the Lorentz factor of the outflow towards the end of the prompt emission or by long lasting 
central engine activity. The decreasing Lorentz factor ($\Gamma$) scenario requires that $E(>\Gamma) \propto \Gamma^{1-s}$ 
with $s>1$, but Nousek et al. find, on the basis of their observed change in decay slope, when 
modelling the light curve with just a single broken power law, that $s = -16.7\pm4.6$ for this burst (see their table $3$), 
thus disallowing this interpretation. However, our more detailed, multi-broken power law analysis shows that this scenario 
is valid for either of our breaks ($s >3$ for both breaks) except when $\nu_{\rm c} < \nu_{\rm x} < \nu_{\rm m}$ for a 
wind medium ($s \sim-21$ and $\sim-63$ for the first and second break respectively).

The long-lasting central engine activity scenario requires that the source luminosity decays slowly with 
time\footnote{$Q$ in the luminosity relation of Nousek et al. (2005) has been capitalised to prevent
confusion with the power law index $q$ used by Panaitescu (2005a)}, 
$L \propto t^{Q}_{\rm lab}$ with $Q>-1$, with the average value found by Nousek et al. being of the order $-0.5$. 
The change in decay slope from their single broken power law model leads the authors to find $Q = 0.3\pm 0.1$ for GRB 050505, 
which is consistent with the lower limit of this mechanism. However, this value of $Q$ is unphysical as it requires the 
luminosity to increase with time. Our analysis shows that the long-lasting central engine activity scenario is valid 
(i.e. $Q<0$, with $Q$ in the range $\sim-0.2$ to $-0.5$), again for either of our breaks, as long as the X-ray frequency, 
$\nu_{\rm x}$, is above the cooling frequency, $\nu_{\rm c}$. We are unable to distinguish, in this case, whether a wind or 
homogenous cirumburst medium is favoured.

Another possible cause of either of the breaks in the light curve of GRB 050505 could be a structured jet outflow. In this case the 
ejecta energy over solid angle, $dE/d\Omega$, is not constant, but varies with the angle $\theta$ measured from the 
outflow symmetry axis~\cite{Meszaros_1998}. Panaitescu~\cite{Panaitescu_2005} suggests that since afterglow light curves 
are power laws in time $dE/d\Omega$ can be approximated as a power law in $\theta$ (see their eqn 13), with a power law 
index of $q$.

We assume a typical value of $p$ (the electron spectral index) to be $2.2$ (Gallant, Achterberg \& Kirk 1999) and use the 
observed values of $\Delta \alpha$ to calculate $q$ from eqns $14$ and $15$ of Panaitescu~\cite{Panaitescu_2005}. This relation only applies 
when $q < \tilde{q}$, where $\tilde{q} = 8/(p+4)$ or $8/(p+3)$. For GRB 050505 the observed values of $\Delta \alpha$ 
give $q$ greater than $\tilde{q}$, within errors, for both wind and uniform environments and for the observing frequency above 
or below the cooling frequency.

For $q > \tilde{q}$, where $dE/d\Omega$ falls off sufficiently fast that the afterglow emission is dominated by the 
core of the jet we expect $\Delta \alpha = 0.75$ (homogenous environment) or $0.5$ (wind environment) (Panaitescu 2005a). 
Thus a structured jet appears to be just consistent with both breaks. However, $\alpha_{1}$ is too shallow to be explained by the 
spherical fireball model, unless the observer is located off the jet core. In this case the value of $\alpha_{1}$ implies 
that our line of sight should be located exceptionally close to the edge of the core.

\begin{figure}
\begin{center}
\includegraphics[width=6cm,angle=270]{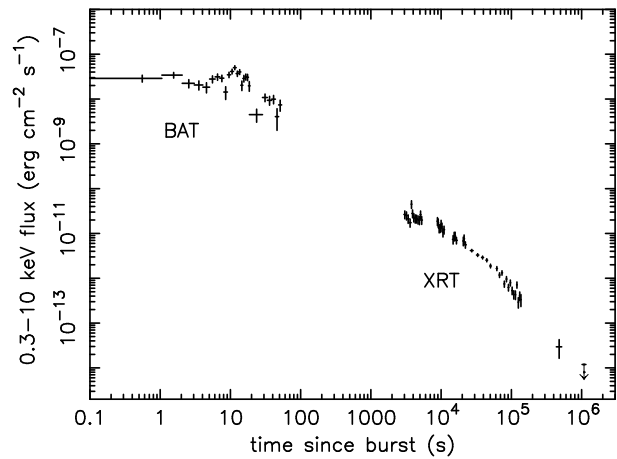}
\caption{The combined BAT-XRT flux light curve, extrapolated into the 
$0.3-10.0$ keV range. For the XRT section of the flux light curve, the countrate was
converted into an unabsorbed flux using the best fit power law model. The
BAT data were extrapolated into the XRT band using the best fit power law model derived from the BAT data alone.}
\end{center}
\end{figure}

The signatures of a jet break, where the relativistic outflow from the GRB slows sufficiently that $\Gamma \sim 1/\theta_{j}$ 
and the jet spreads laterally, are a temporal break with a typical amplitude of $\sim1$~\cite{Rhoads_1999,Chevalier_2000,Sari_1999}, 
no spectral variation~\cite{Piran} and a post-break decay index equal to $p$, the electron spectral index~\cite{Rhoads_1999}. The relation
of $\alpha = p$ post-break is valid for $p > 2$, otherwise a different $\alpha-p$ relation should be adopted~\cite{Zhang_2004,Dai_2001}.
There is no evidence for spectral variation during our observations (see Table 1). Unfortunately there were insufficent optical 
detections of this GRB pre- and post-break to confirm the presence of a jet break in other wavelengths at either epoch.

The temporal index of an X-ray light curve post-jet break should equal $p$, the electron spectral index~\cite{Rhoads_1999}. We 
calculate from our measured spectral index, $\beta$, that  $p = 1.8\pm0.2$ and $2.8\pm0.3$, assuming that $\nu_{\rm x}$ is above and 
below the cooling frequency, $\nu_{\rm c}$, respectively~\cite{Zhang_2004,Sari_1999}. We measure a value of $\alpha_{2} = 1.17^{+0.08}_{-0.09}$,
which is not compatible with either value of $p$, which rules out the first break being due to a jet break. However, 
$\alpha_{3} = 1.97^{+0.27}_{-0.28}$ which agrees, within the limits, to the $\nu_{\rm x} > \nu_{\rm c} $ case ($p = 1.8\pm0.2$). 
However, since $p$ may be $<2$, within the error range, we calculated the expected post-break slope from Dai and Cheng (2001; $\alpha = (p+6)/4$, 
$\nu_{\rm x} > \nu_{\rm c}$) giving an expected decay index of $1.95\pm0.17$, which is also consistent with $\alpha_{3}$. With this value 
of $p$ we can constrain the jet break parameters further~\cite{Rhoads_1999} and conclude that the amplitude of the 
second break is consistent with a value of 0.95, which is the value expected from optically thin synchrotron 
emission when $\nu_{\rm x} > \nu_{\rm c}$, thus supporting the case that the second break is a jet break.  

Having considered the various potential origins for the breaks in the light curve of GRB 050505 for the doubly broken model 
we conclude that the first break is due to the end of energy injection into the forward shock, i.e. that GRB 050505 fits with 
the canonical light curve model proposed by Nousek et al. (2005), and that the second break is due to a jet, either 
structured or uniform. 

The `smoothly broken' core-dominated power law provides a good fit to the XRT light curve data; however, the large degree of smoothing 
involved produces a degeneracy between the smoothing parameter, the first decay index and the break time. If we take 
the example case for $S = 1$ (see Fig $3$), then a break is observed at $T+18.5^{+4.4}_{-3.2}$ ks in the observer's frame.
This translates to $T+3.5^{+0.8}_{-0.6}$ ks in the rest frame of the burst, with $\Delta \alpha = 1.43^{+0.21}_{-0.22}$.

The magnitude of this break allows us to rule out a cooling break or the end of continued energy injection into the forward shock. 
A structured jet could explain the magnitude of this break if the observer is placed off the jet core (Panaitescu 2005b). This would then 
naturally explain the initial shallow decay index and the very smooth break. The magnitude of the break is also compatable with a jet 
break from optically thick synchrotron emission ($\Delta \alpha = 1.25$). However, a break this early requires an unreasonably large 
circumburst density ($n \sim3\times10^{5}$ cm$^{-3}$) to produce a value of $E_{\rm \gamma,rest}$ (Ghirlanda et al. 2004), the true 
$\gamma$-ray energy released, that is comparable with the typical values of $E_{\rm \gamma,rest}$ seen thus far~\cite{Bloom_2003}. 
Thus the parameters of the smoothly broken power law model fit are inconsistent with all of the afterglow models considered here.

\subsection{Multiwavelength Spectral Energy Distribution}

\begin{figure}
\includegraphics[width=\columnwidth]{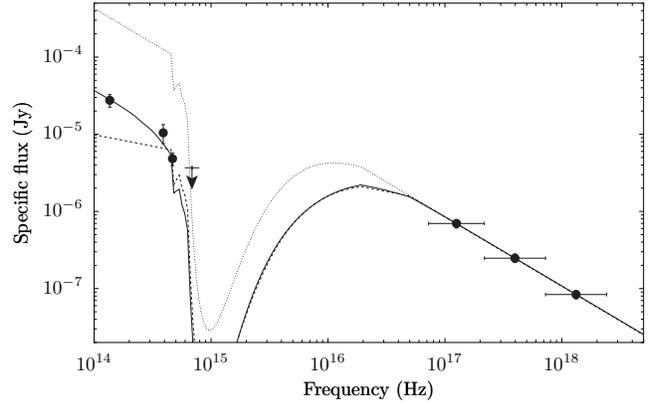}
\caption {\label{figure:bbspectrum}
The optical-nIR to X-ray spectrum of GRB 050505 at $32$ ks after the burst. The X-ray fluxes are corrected for both Galactic and host-galaxy absorption,
while the optical-nIR points have been corrected for Galactic absorption only. The optical R-band point lies on the edge of the Lyman break, with the
Gunn-Peterson trough bluewards of it. The continuous line represents a broken power law, modified by the Lyman break and additional optical/UV host-galaxy
extinction (see text). The dashed line uses the same model, but with no additional extinction. The dotted line is the extrapolation from a single power law
fitted to the X-rays alone, only accounting for the Lyman break.
}
\end{figure}

In Figure \ref{figure:bbspectrum} we show the optical -- X-ray spectrum of GRB 050505. The X-ray fluxes were obtained 
from a spectral fit between $26$ ks and $40$ ks after the bursts; the optical data (UKIRT $K$ band and the FTN data) were scaled 
to a common epoch, chosen to be the logarithmic average of the X-ray data ($32$ ks). The magnitudes have been corrected for 
the estimated Galactic extinction using the dust maps by Schlegel et al. (1998), and have been converted to fluxes using 
the calibration provided by Fukugita et al. (1995) for the optical and that by Tokunaga and Vacca (2005) for the infra red 
magnitudes. Since all optical data were taken between the time of the two breaks, we have used the $\alpha_{2} = 1.17$ light 
curve decay index. However, the decay in the optical can be different. We tested several other values for the decay index 
(at most $0.5$ different from $1.17$), and found the resulting optical fluxes differ at most by $1\sigma$ ($\sim$0.2 mag).

We fit the broad-band spectrum with two basic models, a power law and a broken power law, both accounting for the Lyman break 
(with the redshift fixed at $z = 4.27$) and intrinsic host-galaxy extinction (also with the redshift fixed at $z = 4.27$). 
The Lyman break has been modelled as described in Madau (1995); the optical/UV absorption has been modelled following Pei (1992). 
A single power law is excluded, even allowing for dramatic extinction in the host galaxy ($\chi^{2}/DOF = 
38.19/4$ with the spectral index fixed at $0.9$ as determined from the X-ray data alone). A broken power law, 
with the high-frequency index $\beta_{2}$ also fixed at $0.9$, results in a much better fit. 
We have applied $3$ variants of extinction: none, a Galactic-like extinction curve and an SMC-like extinction curve. The SMC-like 
extinction curve provides a good fit, resulting in the $B-V$ colour excess being $E(B-V) = 0.10 \pm 0.02$ and the low-frequency index
$\beta_{1} = 0.41^{+0.05}_{-0.06}$ ($1\sigma$ confidence limits). The break frequency is largely unrestricted and was kept fixed at a 
value of $10^{16}$ Hz, although values of $10^{17}$ Hz and $10^{15}$ Hz are acceptable (with varying amounts of host-galaxy extinction). 
However, the low number of data points result in a relatively low $\chi^{2}/DOF \sim0.3$, and shows a certain degeneracy: a 
Galactic-like extinction curve results in an equally good fit. This is mostly because the observed wavelength  of the distinct $2175$ \AA\ 
feature\footnote{A strong increase in absorption is found for both the Milky Way and LMC around this wavelength, but is noticeably absent in the SMC (see e.g. Savage \& Mathis, 1979).} falls between our available photometry at this redshift, and the intrinsic extinction is almost 
entirely determined from the 
two $K$ and $I$ band points (the $R$-band point being located on the edge of the Lyman break). The resulting values for a Galactic 
extinction curve are $E(B-V) = 0.20 \pm 0.03$ and $\beta_{1} = 0.50^{+0.06}_{-0.07}$ ($1\sigma$).

The difference between the two power law indices is $\Delta \beta \sim0.5$. To obtain a better constraint for the break frequency, 
we have fixed the indices at $\beta_{1} = 0.4$ and $\beta_{2} = 0.9$. This results in the cooling frequency being located between 
$1.8 \times10^{15}$ Hz and $1.4 \times10^{16}$ Hz (this is dependent on whether a Galactic or SMC extinction curve is used). The 
inferred $E(B-V)$ is the same as before.

Our best fit results favour a cooling break between the optical and X-ray wavebands; in addition, a modest amount of host-galaxy 
extinction would be needed to explain our data fully, but no clear distinction between Galactic or SMC-like extinction can be made. 
A fit with SMC-like extinction, however, agrees marginally better with the expected $\Delta \beta = 0.5$ for a cooling break.

Berger et al. (2005b) measured a Hydrogen column density of $\log N_{\rm HI} = 22.05 \pm 0.10$ from the Ly$\alpha$ absorption in their optical 
spectrum, and a metallicity of $Z \approx 0.06 Z_{\odot}$. We can therefore immediately rule out the Galactic like extinction. Fitting
the X-ray spectrum with intrinsic absorption, setting all elements heavier than He to an abundance of $0.06$, gives 
$N_{\rm H} = 7.43^{+3.77}_{-3.41}\times 10^{22}$ cm$^{-2}$, ie $\log N_{\rm H} = 22.87^{+0.18}_{-0.27}$, in addition to the Galactic absorption
component. This host absorption is higher than the Hydrogen column directly measured by Berger et al. (2005b). It is unlikely that this difference 
is caused by an evolution of the dust and gas properties, since the timescales of the X-ray and optical observations are similar. A reconciliation
of these results can in principle be achieved by ionisation in the host, however, the ionisation fraction required is too high as to be considered 
seriously.

The magnitude of the difference between these two hydrogen column densities is not easily explained. We estimate a 10\% error in the Galactic 
$N_{\rm H}$ in this direction. Setting the Galactic column density to 110\% of its value does not reduce the excess Hydrogen column density in 
the rest frame of the burst sufficiently to reconcile the X-ray absorption with the value of Berger et al. (2005b). Nor can a difference in column 
densities of this magnitude be explained by remaining uncertanties in the XRT calibration.

We also performed a spectral fit allowing both Galactic and host values of $N_{\rm H}$ to vary, rather than constraining the Galactic value to that 
given by Dickey \& Lockman (1990), using the XSPEC STEPPAR command to explore the absorption column parameter space. The host absorption column still 
exceeded the value given by Berger et al. (2005b) at greater than 90\% confidence. We speculate that that some curvature of unknown origin may be 
present in the X-ray spectrum.

From the Hydrogen column density, and using the relation between $N_{\rm HI}$ and $E(B-V)$ for the SMC (Martin et al. 1989), we can infer $E(B-V) = 0.24$. 
We note that this value is likely to be lower, with the metallicity
being half of the estimated SMC ISM metallicity (Pei 1992). The inferred value is moderately in agreement with the $E(B-V) = 0.10$ we find from
directly fitting the optical -- X-ray spectrum with an SMC-like extinction curve (assuming $R_{V} = 2.39$), although the Galactic
extinction curve results in an extinction measurement which is equally well compatible with the inferred $E(B-V)$. This approximately agrees with 
$A_{V} = R_{V}\cdot E(B-V) = 0.3$ as found by Berger et al. (2005b). Such a low extinction
value is not uncommonly seen in GRB afterglows (e.g. Galama \& Wijers 2001, Stratta et al. 2004).

\subsection{Burst Properties}

From the redshift of GRB 050505 ($z = 4.27$) and the mean flux over the observed $15-350$ keV $T_{90}$ spectrum we calculate an isotropic equivalent 
radiated energy, $E_{\rm iso,rest}$, in the extrapolated $1-10^{4}$ keV rest frame energy range to be $4.44^{+0.80}_{-1.12}$$\times10^{53}$ ergs, 
using the standard cosmology~\cite{Spergel_2003}: $H_{\rm o} = 71$ km s$^{-1}$ Mpc$^{-1}$, ($\Omega_{M}, \Omega_{\Lambda}$) 
$=$ ($0.27, 0.73$), and a K-correction of $3.09^{+0.48}_{-0.33}$.

If we take the second break in the light curve to be a jet break we are then able to calculate the properties of 
GRB 050505. Using the formulation of Frail et al.~\cite{Frail_2001}, and assuming that the efficiency of the fireball in converting 
the energy of the ejecta into $\gamma$-rays is $\sim0.2$, we obtain a range in $\theta_{\rm j}$ from $2.2^{\rm o}$ ($n = 
1$ cm$^{-3}$) to $3.8^{\rm o}$ ($n = 100$ cm$^{-3}$) ~\cite{Panaitescu}. Frail et al. (2001) conclude that opening angles of $\leq3^{o}$
are required for less than 10 per cent of the $BeppoSAX$ GRB sample. However, such a narrow beaming angle would not be unexpected for a 
high redshift burst as GRBs with wide jets would be too faint to be detected by current $\gamma$-ray missions.

From this we can calculate the beaming fraction $f_{b}=(1-\cos\theta_{\rm j})$ ~\cite{Sari_1999} to be between 
$7.1\times10^{-4}$ ($n = 1$ cm$^{-3}$) and $2.3\times10^{-3}$ ($n = 100$ cm$^{-3}$) and $E_{\rm \gamma,rest}$, the true 
$\gamma$-ray energy released, to be in the range of $3.17^{+0.86}_{-1.11} \times10^{50}$ ($n = 1$ cm$^{-3}$) to 
$9.99^{+3.00}_{-3.24} \times10^{50}$ ergs ($n = 100$ cm$^{-3}$) for a rest frame energy band of $1-10^{4}$ keV. 
We note that the typical $E_{\rm \gamma,rest}$ of bursts thus far is $9.8\times10^{50}$ ergs ~\cite{Bloom_2003} 
with a burst-to-burst variance about this value of $\sim0.35$ dex (or a factor of $2.2$), thus this burst agrees 
well with the typical value provided the circumburst density is of the order $100$ cm$^{-3}$.

We found it useful to calculate $E_{\rm peak,rest}$ from these values of $E_{\rm \gamma,rest}$ via the Ghirlanda relation 
~\cite{Ghirlanda_2004} and compare these values to the observed lower limit of $E_{\rm peak,obs}$ $>52$ keV 
($E_{\rm peak,rest} > 274$ keV). We calculated that the Ghirlanda relation gave $E_{\rm peak,rest}$ = $215^{+39}_{-51}$ keV 
(for $n=1$ cm$^{-3}$) and $484^{+130}_{-125}$ keV (for $n=100$ cm$^{-3}$), which agrees with the lower observed limit if 
the circumburst density is high. We also calculated $E_{\rm peak,rest}$ via the Amati correlation~\cite{Amati_2002,Lloyd_2002}. 
Using equation $6$ of Ghirlanda et al.~\cite{Ghirlanda_2005} for GRBs of known redshift gives $E_{\rm peak,rest} = 1000^{+115}_{-151}$ keV, 
consistent with our observed limit.

\section{Conclusions}

We have presented multi-wavelength data for GRB 050505. Our earliest X-ray data starts $\sim47$ minutes after the 
GRB trigger time as the {\it Swift} satellite was unable to slew to it immediately due to an Earth limb constraint. 
 The X-ray light curve of GRB 050505 (see Figs $2$ and $3$) can be adequately fit with either a `smoothly broken' or
`doubly broken' power law model.

The `smoothly broken' power law model (see Fig $3$) favours a smoothing factor of $0.5-2$ (highly smoothed transition). This 
produces an initially shallow decay with $\alpha_{1} \sim0.5$, which breaks over several decades in time to a 
steeper slope, $\alpha_{2}$, of $\sim1.8$. ($\chi^2/DOF \sim1.04$) The values of the decay indices are poorly
constrained but, assuming a smoothing parameter $S = 1$, then a break is observed at $T+18.5^{+4.4}_{-3.2}$ ks in the observer's 
frame with $\Delta \alpha = 1.43^{+0.21}_{-0.22}$. The magnitude of this break is inconsistent with all of the afterglow models
considered here.

A `doubly broken' power law model (see Fig $2$) consists of a shallow decay, $\alpha_{1} = 0.25^{+0.16}_{-0.17}$, first detected 
at $T+3$ ks, followed by a break in the observer's frame at $t_{1} = 7.4^{+1.5}_{-1.5}$ ks and a steeper 
decay $\alpha_{2} = 1.17^{+0.08}_{-0.09}$. This decay breaks sharply again at $t_{2} = 58^{+9.9}_{-15.4}$ ks into a 
yet more rapidly decaying index of $\alpha_{3}$ = $1.97^{+0.27}_{-0.28}$, which continues until at least 
$T+\sim500$ ks ($\chi^{2}/DOF = 38.7/42$).

We see no change in the X-ray spectral properties during {\it Swift's\/} observations of this GRB. The best fit model 
parameters for the X-ray spectrum indicates that this burst has a typical photon index of $1.90^{+0.08}_{-0.08}$ and 
an excess absorption component from the host galaxy of ($1.28^{+0.61}_{-0.58}$)$\times10^{22} \rm cm^{-2}$ 
($\chi^{2}/DOF = 99/96$). 

Having considered the temporal position and amplitude of the two breaks in the doubly broken light curve we conclude that the 
first break is due to the end of energy injection into the forward shock~\cite{Nousek_2005,Zhang_2005}, i.e. that 
GRB 050505 fits with the canonical light curve model proposed by Nousek et al. (2005), and that the second break 
is jet break caused by either a structured or uniform jet.

The optical -- X-ray spectrum indicates that the cooling break is located between the optical and X-ray bands, as seen in many other GRB
afterglows. A modest amount of intrinsic UV/optical extinction is required in addition, which for an SMC-like extinction law
would result in $E(B-V) = 0.10$. We note that a Galactic extinction law fits equally well, but the $0.06$ Solar metallicity inferred from
the optical spectrum (Berger et al 2005b) shows it to be more SMC-like. Interestingly, the $N_{\rm H}$ column density
inferred from the X-ray spectrum with the metallicity set to $0.06 Z_{\odot}$ is higher than that directly measured from the HI
column.

The redshift of $4.27$ allowed us to calculate the intrinsic parameters for this GRB, in conjunction with the second
light curve break time observed in {\it Swift's} X-ray observations. The identification of this break with a jet break 
provides a value for $E_{\rm \gamma,rest}$ that is in good agreement with respect to previous GRBs, provided that the 
circumburst density is of the order $100$ cm$^{-3}$ and the values are consistent with the Ghirlanda~\cite{Ghirlanda_2004,Ghirlanda_2005}
and Amati~\cite{Amati_2002,Lloyd_2002} relations. It also suggests that GRB 050505 has a narrow beaming angle; 
however, this degree of beaming is not unexpected for GRBs at high redshift since GRBs with wider jets could potentially 
be too faint to be detected by any of the current $\gamma$-ray missions.

\section*{Acknowledgments}

This work is supported at the University of Leicester by the Particle Physics and Astronomy 
Research Council (PPARC) and at Penn State by NASA contract NAS5-00136. The Faulkes Telescope 
North is supported by the Dill Faulkes Educational Trust. CPH gratefully acknowledges support 
from a PPARC studentship. VLP, MP and PR gratefully acknowledge ASI grant I/R/039/04. The 
authors acknowledge Iain Steele (LJMU) and Cristiano Guidorzi (LJMU) for their assistence 
with the FTN observations and Rhaana Starling (Univeristy of Amsterdam) for useful discussions.
We are also grateful to the referee, Bruce Gendre, for his constructive and valuable comments.

\bsp

\label{lastpage}

\end{document}